\documentclass[letterpaper]{article} 
\usepackage{aaai2026}  
\usepackage{times}  
\usepackage{helvet}  
\usepackage{courier}  
\usepackage[hyphens]{url}  
\usepackage{graphicx} 
\usepackage{natbib}  
\usepackage{caption} 
\usepackage{algorithm}
\usepackage{algorithmic}

\usepackage{xcolor} 
\usepackage{subfigure}
\usepackage{graphicx}
\usepackage{color}
\usepackage{booktabs}
\usepackage{mathrsfs}
\usepackage{amsmath} 
\usepackage{amsfonts}
\usepackage{tikz}
\usepackage{array}
\usepackage{multirow}
\usepackage{makecell}
\usepackage[utf8]{inputenc}
\usepackage{amssymb}

\usepackage{newfloat}
\usepackage{listings}
\DeclareCaptionStyle{ruled}{labelfont=normalfont,labelsep=colon,strut=off} 
\floatstyle{ruled}
\newfloat{listing}{tb}{lst}{}
\floatname{listing}{Listing}
\title{CroPS: Improving Dense Retrieval with Cross-Perspective Positive Samples in Short-Video Search}
\author{
    Ao Xie\textsuperscript{\rm 1}\thanks{Contribution during internship at Kuaishou Technology.}\equalcontrib, 
    Jiahui Chen\textsuperscript{\rm 1}\equalcontrib, 
    Quanzhi Zhu\textsuperscript{\rm 1}\equalcontrib, 
    Xiaoze Jiang\textsuperscript{\rm 1}\thanks{Corresponding author.}, 
    Zhiheng Qin\textsuperscript{\rm 1}, 
    Enyun Yu\textsuperscript{\rm 1}, 
    Han Li\textsuperscript{\rm 1}
}
\affiliations{
    \textsuperscript{\rm 1}Kuaishou Technology, Beijing, China\\

    \{xieao03, chenjiahui11, zhuquanzhi03, jiangxiaoze, lihan08\}@kuaishou.com, \\ qinzhiheng1991@gmail.com, yuenyun@126.com
}

\usepackage{bibentry}

\begin{document}

\maketitle

\begin{abstract}
Dense retrieval has become a foundational paradigm in modern search systems, especially on short-video platforms. However, most industrial systems adopt a self-reinforcing training pipeline that relies on historically exposed user interactions for supervision. This paradigm inevitably leads to a filter bubble effect, where potentially relevant but previously unseen content is excluded from the training signal, biasing the model toward narrow and conservative retrieval. In this paper, we present CroPS (Cross-Perspective Positive Samples), a novel retrieval data engine designed to alleviate this problem by introducing diverse and semantically meaningful positive examples from multiple perspectives. CroPS enhances training with positive signals derived from user query reformulation behavior (query-level), engagement data in recommendation streams (system-level), and world knowledge synthesized by large language models (knowledge-level). To effectively utilize these heterogeneous signals, we introduce a Hierarchical Label Assignment (HLA) strategy and a corresponding H-InfoNCE loss that together enable fine-grained, relevance-aware optimization. Extensive experiments conducted on Kuaishou Search, a large-scale commercial short-video search platform, demonstrate that CroPS significantly outperforms strong baselines both offline and in live A/B tests, achieving superior retrieval performance and reducing query reformulation rates. CroPS is now fully deployed in Kuaishou Search, serving hundreds of millions of users daily.
\end{abstract}

\section{Introduction}
\begin{figure}[t]
    \centering
    \includegraphics[width=\linewidth, trim=285 170 285 150, clip]{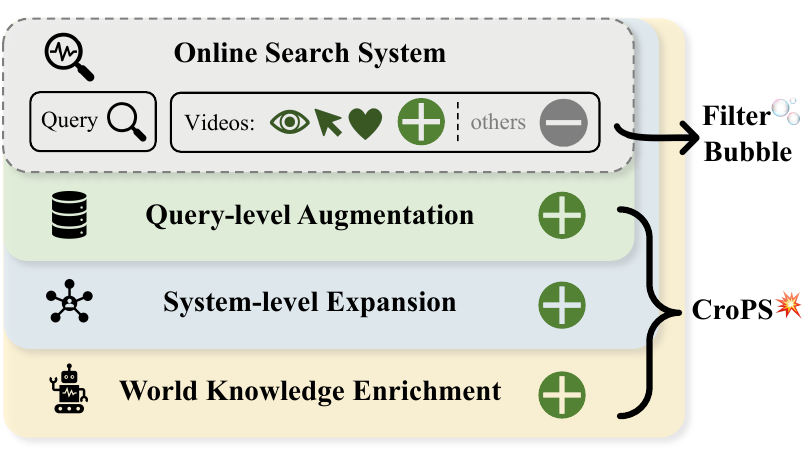}
    \caption{Overview of CroPS, where ``+" (``-")  represents positives (negatives). It introduces positives from three complementary sources: Query-level Augmentation, System-level Expansion, and World Knowledge Enrichment.}
    \label{fig:teaser}
\end{figure}

Dense retrieval has emerged as a foundational paradigm in modern search systems due to its efficiency and effectiveness in matching semantically rich queries to relevant documents~\cite{zhang2022multi,ma2024fine}. In short-video search platforms, dual-encoder retrievers (often referred to as “two-tower” models) are widely adopted, where separate encoders are used to independently embed user queries and candidate videos into a shared latent space. 
Typically, these retrievers are trained using a data-driven and self-reinforcing paradigm: as shown in Figure~\ref{fig:teaser} (gray block), historical user interactions (\textit{e.g.}, watching, clicking, or liking) from a production search system serve as supervision. Videos previously displayed for a given query are labeled as positive examples, while negative samples are drawn from content that was filtered out during earlier retrieval stages or never exposed to the user~\cite{zheng2024full}. While this paradigm facilitates iterative model improvement through continuous data accumulation, it also poses a critical challenge: the filter bubble effect. 
Because only historically exposed content is considered as positive supervision, potentially relevant but previously unseen videos are excluded from the positive set, even incorrectly labeled as negatives. This bias restricts the model’s ability to surface novel or diverse content, leading to conservative and narrow retrieval behavior. The resulting lack of diversity impairs user experience and limits the system’s capacity to support exploratory search. 

Previous research has primarily focused on improving retrieval performance through architectural innovations \cite{zhang2022multi,khattab2020colbert} or refined negative sampling strategies \cite{karpukhin2020dense,yang2024trisampler}. However, these efforts remain constrained by the limitations of the self-reinforcing training pipeline and thus cannot fully escape the filter bubble \cite{meghwani2025hard,ren2021pair}. In contrast, we propose that positive sample enrichment offers a powerful and complementary solution that has been largely overlooked. By introducing relevant content beyond the scope of historical exposure, we can effectively break through the boundaries of existing data.

In this paper, we propose CroPS (Cross-Perspective Positive Samples), a retrieval enhancement data engine that incorporates diverse positive signals from multiple perspectives to break the limitations imposed by the filter bubble (see Figure~\ref{fig:teaser}). CroPS enriches the training signal with previously unobserved yet semantically relevant examples from three complementary sources. First, it captures \textbf{query-level augmentation} by identifying the user query reformulation behavior. When users issue a follow-up query with similar semantics after an unsatisfactory search, CroPS treats videos consumed after such reformulations as additional positives for the original query. This leverages intent continuity to expand the positive set beyond what the system originally retrieved. Second, CroPS introduces \textbf{system-level augmentation} by bridging search and recommendation systems. It incorporates videos from the recommendation flow that align semantically with the user’s search query, thus breaking the silo between independently functioning subsystems and leveraging user engagement data across systems. Finally, CroPS performs \textbf{world knowledge augmentation} by using large language models as virtual retrievers. These models, endowed with rich semantic and factual knowledge, are used to synthesize high-quality training signals that reflect relevant content beyond what is observable in user interaction logs. Meanwhile, it  simulates user behavior where, upon failing to obtain satisfactory information within the current app, users seek knowledge from alternative sources. This is regarded as a form of cross-platform knowledge integration, which can effectively mitigate the effects of information cocoons. Together, these augmentations significantly expand the positive space, allowing the retriever to learn a more comprehensive representation of relevance and generalize beyond historically exposed content.

However, naively treating all these heterogeneous positives as equal during training leads to suboptimal learning. We propose a Hierarchical Label Assignment (HLA) strategy that assigns different levels to positive samples based on their origin and reliability. This hierarchical supervision allows the model to learn fine-grained ranking behaviors aligned with the system objectives. For instance, by assigning higher labels to query-level augment positives, the model learns to generalize across reformulated queries, potentially reducing the frequency of user query rewrites. To support HLA-based training, we introduce H-InfoNCE, a novel loss function tailored to hierarchical supervision, enabling more targeted optimization across different levels of relevance. Unlike InfoNCE~\cite{oord2018representation}, which individually compares each positive sample with its corresponding negatives, H-InfoNCE requires that all positive samples from higher tiers be jointly contrasted with negatives from lower tiers in a single step. By leveraging efficient and large-scale positive-negative contrastive learning, H-InfoNCE empowers our model to achieve rapid and effective learning, leading to notable performance gains and providing meaningful insights for practical industrial applications.

In summary, this work makes three key contributions.
(1) We identify a fundamental limitation in industrial dense retrieval systems: the presence of filter bubbles caused by self-reinforcing training paradigms, and argue that introducing positive samples beyond the historical exposure space is an effective and underexplored solution. 
(2) We propose CroPS, a novel framework that enriches the positive training set by incorporating semantically relevant examples from query reformulations, cross-system user feedback, and external world knowledge via large language models. To support learning from this heterogeneous supervision, we introduce a Hierarchical Label Assignment (HLA) strategy along with a tailored H-InfoNCE loss, enabling the model to capture fine-grained ranking preferences aligned with system-level goals. 
(3) We conduct extensive offline and online evaluations on Kuaishou Search, a large-scale short-video search platform. The results demonstrate that CroPS not only achieves state-of-the-art retrieval performance in offline experiments, but also significantly improves user engagement and reduces query reformulation rates in online A/B testing. CroPS has been successfully deployed in Kuaishou Search and now serves as a core component of the search infrastructure for hundreds of millions of active users. 

\section{Related Works}

\noindent \textbf{Dense Retrieval.} 
The modern model commonly employs a dual-encoder architecture, which is well adapted to approximate nearest neighbor (ANN) \cite{johnson2021billion} retrieval. Methods in dense retrieval can generally be categorized into five main areas: retrieval-oriented continued pre-training \cite{su2024wikiformer,oguz2022domain,chang2020pre}, negative sampling strategies \cite{meghwani2025hard,yang2024trisampler,zhou2022simans}, structural augmentation techniques \cite{luan2021sparse,khattab2020colbert,humeau2020poly}, the exploration of generative paradigms \cite{li2025matching,li2024unigen,tang2024list,wang2025personalized}, and the applications in specific industrial scenarios \cite{lin2024enhancing,rossi2024relevance,zheng2022multi}. These works have not conducted an in-depth study of positive samples and failed to address the information cocoon effect in the retrieval domain. Furthermore, the reliance on complex post-interaction models like the Poly-encoder \cite{humeau2020poly} presents significant challenges for integration with ANN-based systems, which in turn hampers practical deployment. Unlike previous works, our architecture-agnostic approach can be applied to enhance any model using CroPS without introducing any inference overhead, making it well-suited for online applications. At the same time, it mitigates the information cocoon effect in retrieval systems by leveraging positive sample enhancement and a hierarchically stratified contrastive loss, thereby increasing the diversity of search results and improving the overall user experience.

\begin{figure*}[t]
\centering
\includegraphics[width=1.0\textwidth, trim=20 200 20 200, clip]{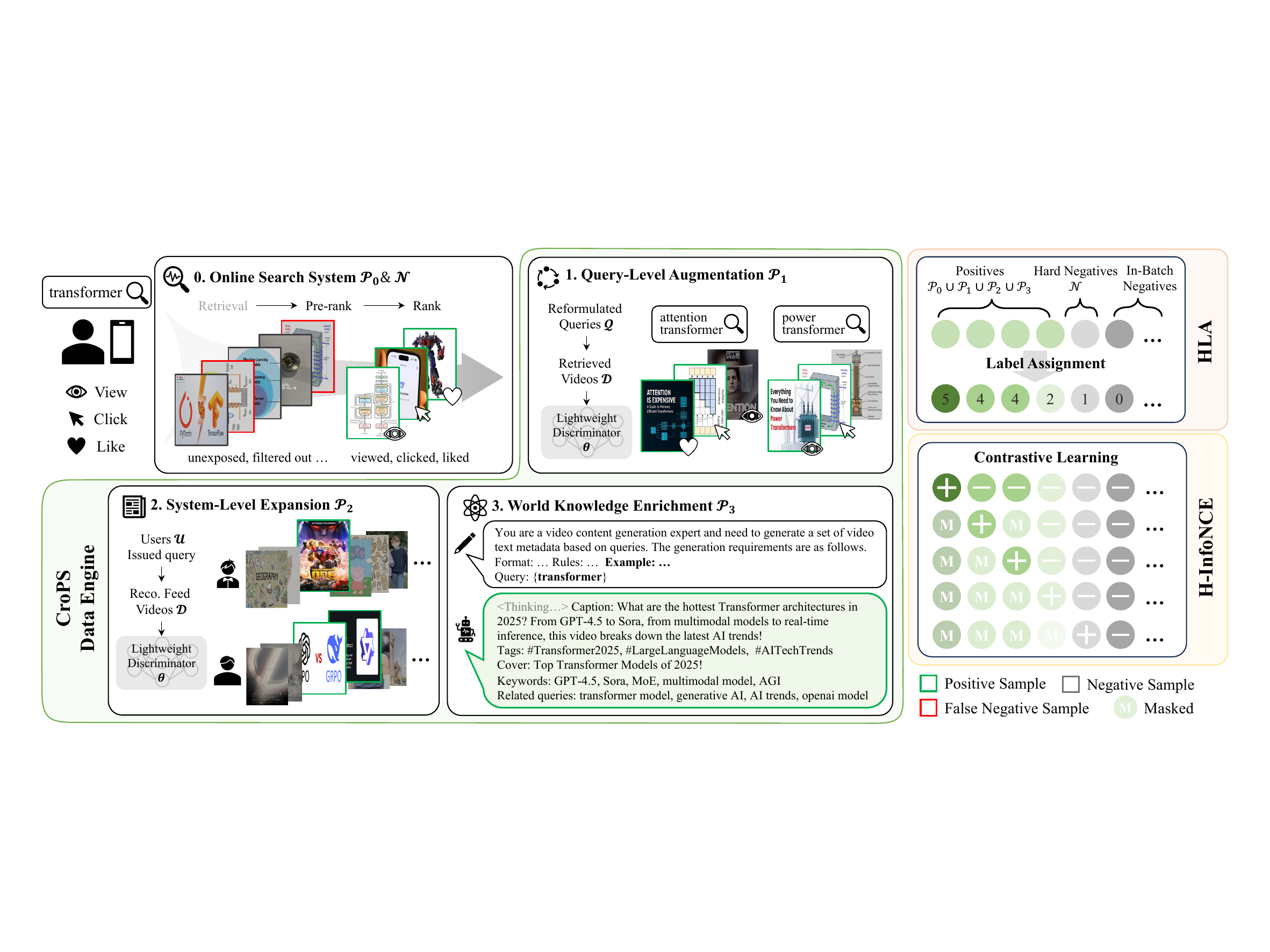}
\caption{The overall framework of the CroPS, where ``+" (``-") represents positives (negatives). It consists of three components: CroPS Data Engine, Hierarchical Label Assignment (HLA), and H-InfoNCE Optimization. CroPS Data Engine expands the positive samples from: (1) Query-level Augmentation, (2) System-level Expansion, and (3) World Knowledge Enrichment.}
\label{fig:crops}
\end{figure*}

\noindent \textbf{Contrastive Learning.}
Contrastive learning has become a fundamental paradigm for representation learning in both vision and language domains~\cite{wu2018unsupervised,he2020momentum,radford2021learning,jiang2022xlm}. In retrieval, contrastive objectives such as InfoNCE~\cite{oord2018representation} and its variants~\cite{li2021embedding,magnani2022semantic,hoffmann2022ranking} form the backbone of training strategies by maximizing the similarity between query-positive pairs while pushing apart query-negative pairs. Various works have extended this framework to enhance efficiency and robustness, including hard negative mining~\cite{yang2024trisampler}, in-batch negatives~\cite{radford2021learning}, and list-wise formulations~\cite{hoffmann2022ranking}. However, most of these approaches assume a flat binary relevance structure and fail to capture the nuanced semantic relationships between different types of positive and negative samples.
Recent efforts have explored hierarchical or weighted contrastive learning, yet these typically rely on pre-defined heuristics or coarse metadata for label assignment. In contrast, our proposed HLA (Hierarchical Label Assignment) introduces a data-driven, semantically grounded label hierarchy over both positive and negative samples. Built upon this, we propose H-InfoNCE, a stratified contrastive loss that enables the model to learn graded relevance and fine-grained ranking preferences. It better reflects real-world retrieval scenarios, where training data often contain heterogeneous supervision signals with varying confidence levels.

\section{Methodology}

In industrial short-video search and recommendation, textual information—comprising user-edited titles, captions, keywords, and OCR-extracted text fields—offers an effective, cost-efficient video representation \cite{zheng2024full,zheng2022multi}. As a standard practice, our retrieval model also utilizes this textual information as input to encode videos.

\subsection{CroPS Data Engine}
Different from prior works that primarily focus on better negative sampling strategies for retriever training \cite{meghwani2025hard,yang2024trisampler}, CroPS takes a complementary perspective by enriching the positive sample set. By incorporating multiple sources of semantically relevant items, it breaks through the limitations of the conventional feedback loop and mitigates the filter bubble effect. 

As shown in Figure \ref{fig:crops}, given a query $q$, the CroPS data engine collects the positive set $\mathcal{P} = \mathcal{P}_{0} \cup \mathcal{P}_{1} \cup \mathcal{P}_{2} \cup \mathcal{P}_{3} = \{d^{+}_{1}, d^{+}_{1}, d^{+}_{2}, ..., d^{+}_{k}\}$ from the four perspectives below, and the negative set $\mathcal{N} = \{d^{-}_{k + 1}, d^{-}_{k + 2}, ..., d^{-}_{n}\}$.

\noindent \textbf{Online Search System. } 
The retrieval process in short-video search systems typically follows a multi-stage pipeline consisting of recall, pre-ranking, and ranking, which sequentially filters and ranks candidate videos. Following common practice~\cite{zheng2024full}, we treat videos that were clicked or liked during the ranking stage as a basic positive set $\mathcal{P}_{0}$. For the negative sample set $\mathcal{N}$, we sample hard negatives from two sources: videos that were not exposed to users in the ranking stage, and those that filtered out between the pre-ranking and ranking stages. This sampling strategy strikes a balance between positive relevance and negative difficulty, ensuring a meaningful training signal. 
However, relying solely on interaction-based labels can reinforce filter bubble effects: relevant but previously unexposed content is systematically excluded from positive training, introducing bias and limiting the model’s ability to surface novel but pertinent results. 
As illustrated in Figure~\ref{fig:crops}, for the query ``transformer'', user interactions are dominated by videos about transformer models in deep learning or the Transformers franchise, while videos related to electrical transformers are rarely exposed and thus incorrectly categorized as negatives, despite being semantically relevant. To mitigate such false negatives, CroPS enriches the positive set with diverse and semantically relevant samples, promoting broader content coverage and improving retrieval robustness.

\noindent \textbf{Query-Level Positive Augmentation. }
Query reformulation is a common user behavior in search systems, often indicating a continued effort to refine or clarify an underlying information need. When a user issues a semantically related follow-up query shortly after an initial one, the videos consumed in response to the new query are likely to remain relevant to the original intent, even if they were not retrieved in the initial search. These samples typically lie beyond the retrieval scope of the original query and thus offer valuable signals for expanding the positive set, enhancing both recall diversity and model generalization.
To identify such samples, we employ a lightweight semantic discriminator $\theta(\cdot)$ to measure the relevance between the original query $q$ and candidate videos $d_{ij}$ consumed after reformulated queries. The query-level positive set is defined as:
\begin{equation}
\mathcal{P}_{1} = \bigcup_{q_i \in \mathcal{Q}} \{ d_{ij} \in \mathcal{D}_{i} \mid \theta(q, d_{ij}) > \alpha \}
\label{eq:query_level_aug}
\end{equation}
Here, $q$ is the original query, $\mathcal{Q}$ is the set of reformulated queries issued by the same user within a short time window (\textit{e.g.}, 90 seconds), $\mathcal{D}_i$ is the set of videos the user interacted with (e.g., clicked or watched) after issuing $q_i \in \mathcal{Q}$, and $\alpha$ is a predefined semantic relevance threshold.
As illustrated in Figure~\ref{fig:crops}, consider a user initially searching for ``transformer'' but receiving unsatisfactory results. They may reformulate the query to ``power transformer'' to clarify the intent. The relevant videos retrieved and consumed under the reformulated query can serve as positive samples for the original query ``transformer''. This approach not only corrects potential false negatives in $\mathcal{N}$ but also reduces the overemphasis on dominant interpretations, thus promoting more balanced and accurate retrieval.

\noindent \textbf{System-Level Positive Expansion.}
In addition to the search system, short-video platforms operate a recommendation system that independently serves a large portion of user traffic. The recommendation feed typically generates extensive user interaction data, which contains valuable signals that can help mitigate the information cocoon effect. CroPS leverages this opportunity by bridging the gap between search and recommendation, enabling cross-system positive signal integration to further enhance training supervision.
Specifically, for a given query $q$, we identify a set of users $\mathcal{U}$ who have issued this query and retrieve up to 100 videos $\mathcal{D}i$ that each user $u_i \in \mathcal{U}$ interacted with from the recommendation feed, within a temporal window centered around the query timestamp. A lightweight semantic discriminator $\theta(\cdot)$ is employed to evaluate the relevance between the query $q$ and each candidate video $d_{ij}$. Those with scores exceeding a threshold $\alpha$ are included in the system-level positive set, formally defined as:
\begin{equation}
\mathcal{P}_{2} = \bigcup_{u_i \in \mathcal{U}} \{ d_{ij} \in \mathcal{D}_{i} \mid \theta(q, d_{ij}) > \alpha \}
\label{eq:eq:system_level_aug}
\end{equation}
This strategy enables the model to incorporate semantically relevant signals that are not captured by the search pipeline, thereby increasing the coverage of positive examples and mitigating exposure bias. Compared to search-derived samples, recommendation-based positives are typically more timely and better aligned with users' personal interests, offering a valuable supplement to query-based supervision.

\noindent \textbf{World Knowledge Enrichment via LLMs.}
Despite the incorporation of query-level and system-level positive sample enrichment, the filter bubble effect may still persist due to the inherent limitations of existing search and recommendation engines, as well as the finite scope of the platform’s video inventory. To further mitigate this constraint and break through the platform’s content exposure boundaries, we introduce a world knowledge enrichment strategy that leverages large language models (LLMs) as an external knowledge source.
Specifically, the LLM is employed as a pseudo-retriever to synthesize high-quality positive examples $\mathcal{P}_3$. As shown in Figure~\ref{fig:crops}, we adopt a one-shot prompting strategy: the LLM is provided with a query and an exemplar video deemed relevant, and is then instructed to generate other video descriptions aligned with the query. These synthetic examples serve as positive samples that enrich the training set with diverse linguistic expressions, conceptual variations, and external knowledge not represented in the platform’s existing content or logs.
Importantly, these generations capture latent semantic associations and factual information from the broader world, enabling the retrieval model to generalize beyond the scope of historical user interactions. By integrating LLM-generated positives into the training process, we expand the model’s relevance coverage, enhance its capacity to surface novel but meaningful content, and further reduce the impact of content exposure bias.

\begin{table}[t]
\centering
\small
\renewcommand{\arraystretch}{1.1}
\begin{tabular}{c|l|c}
\toprule
     & \textbf{Sample Types} & \textbf{Label} \\
\midrule
 \textbf{Positive} & Query-level aug. positives  & 5 \\
 \multirow{6}{*}{%
    \begin{tikzpicture}[baseline=(current bounding box.center)]
        \draw[->, thick] (0.0,1.0) -- (0.0,-1.0);
    \end{tikzpicture}
}
                & System-level aug. positives             & 4 \\
                & World-knowledge aug. positives                  & 4 \\
                & Clicked videos in rank                & 4 \\
                & Exposed videos in rank                  & 3 \\
                & Unexposed videos in rank  & 2 \\
                & Filtered between pre-rank and rank & 1 \\
\textbf{Negative} & In-batch negatives & 0 \\
\bottomrule
\end{tabular}
\caption{Hierarchical Label Assignment (HLA) for positive and negative samples.}
\label{tab.hla}
\end{table}

\subsection{Hierarchical Label Assignment}
The CroPS data engine introduces positive samples from multiple perspectives. However, these samples vary in importance and reliability. Treating them uniformly can lead to suboptimal learning outcomes. To address this, we propose a Hierarchical Label Assignment (HLA) mechanism. Rather than assigning equal importance to all positive examples, HLA allocates discrete label levels ranging from 0 to 5 \footnote{The label range is flexible; only the relative order is essential.} to reflect varying degrees of relevance and to guide the retriever’s ranking preferences accordingly. This hierarchical supervision enables the model to learn fine-grained semantic distinctions and supports flexible optimization for different training objectives.

Formally, for a query $q$ and its sample set $\mathcal{S} = \mathcal{P} \cup \mathcal{N} = \{d_{1}, d_{2}, ..., d_{n}\}$ that contains $n$ positive/negative videos, HLA assigns $n$ labels $\{l_{ 1}, l_{2}, ..., l_{n}\}$ to each sample $d_i$ based on its origin, as detailed in Table~\ref{tab.hla}. 
Query-level augmented positives receive the highest label (5) because they most directly reflect refined user intent: the reformulated queries are typically issued by users when initial results are unsatisfactory, and the subsequent interactions often provide highly relevant signals. A label of 4 is given to system-level and world-knowledge augmented positives, as well as clicked videos, all of which provide strong relevance signals from different perspectives. Exposed but unclicked videos in the ranking stage are labeled 3, indicating moderate relevance. Unexposed videos in the ranker’s output receive label 2, capturing weak or uncertain relevance. Videos filtered between pre-rank and rank are assigned label 1, while in-batch negatives receive the lowest label of 0. 
This hierarchical scheme enables the retriever to exploit richer supervision beyond binary feedback and learn to rank content in accordance with nuanced relevance signals, ultimately improving both recall diversity and user satisfaction.

\subsection{H-InfoNCE Optimization}

To effectively leverage the multi-grade supervision provided by HLA, we adopt the Hierarchical InfoNCE (H-InfoNCE) loss function. Unlike the standard InfoNCE loss, which assumes binary relevance, H-InfoNCE imposes a label-aware contrastive structure: for a given positive sample with label $l$, only samples with strictly lower labels ($< l$) are considered as negatives. This formulation, shown in Eq.~\ref{eq:eq:h-infonce}, better aligns with the graded supervision and guides the model to learn fine-grained, list-wise ranking preferences.
\begin{small}
\begin{equation}
    \mathcal{L} = 
    -\sum_{d_i\in \mathcal{S}}\log \frac{ \exp\left(\frac{\mathrm{sim}(q, d_i)}{\tau}\right) }{ \sum_{d_j\in \{d_i\} \cup \{d_k \in S | l_i>l_k\}} \exp\left(\frac{\mathrm{sim}(q, d_j)}{\tau}\right) }
\label{eq:eq:h-infonce}
\end{equation}
\end{small}

\noindent We ensure training scalability by efficiently implementing H-InfoNCE, using a mask matrix to filter non-comparable samples by label (see Figure~\ref{fig:crops}, where samples with labels $\geq l$ are masked) and a label-indexed data structure to organize inputs. This allows computation of all hierarchical contrastive losses in one forward pass, matching standard InfoNCE speed and making it suitable for industrial-scale deployment for large-scale data. More details can be found in the Appendix.

\section{Experiments}

\subsection{Dataset, Metrics and Details}

\noindent {\bf Dataset.} Public datasets for information retrieval evaluation often fail to meet the demands of real-world applications. These datasets only record single search queries, ignoring the complex user behavior patterns found in actual systems. Thus, we create a new dataset, called CPSQA, that captures diverse user behaviors, including query reformulation sequences and cross-domain consumption patterns covering search and recommendation systems. 
CPSQA is developed using real user interaction logs from Kuaishou Search, capturing authentic user behavior patterns at an industrial scale. Specifically, we utilize logs from June 11, 2025, for training (500 million samples) and logs from June 13, 2025, for testing (10,000 samples). The CPSQA reflects the scale of online production environments, incorporating a candidate item pool of about 1 billion items. 

\noindent {\bf Metrics.} For evaluation, we employ Recall@100 on two test sets: {\it click-through data} (CT) and {\it query reformulation data} (QR). The CT test set treats clicked videos as positives, demonstrating immediate user preferences. The QR test set comprises videos consumed following query reformulation, which are validated by a semantic discriminator based on a predefined relevance threshold. It is designed to evaluate the model’s ability to generalize beyond historical exposure patterns and mitigate the filter bubble effect. In addition, we conduct NDCG@4 evaluation on the manually annotated test  dataset. NDCG@4 is utilized to assess the recall model's ability to rank the top 4 results, serving only as a reference monitoring metric. The primary evaluation metric for the recall model is still the recall rate.

\noindent \textbf{Implementation Details.} The CroPS data engine comprises three key components.
For query-level positive sample augmentation, we identify user query reformulation behavior by detecting pairs of semantically similar queries issued by the same user within a 90-second window. The lightweight discriminator $\theta(\cdot)$ that we used is a pre-trained 6-layer Transformer model. The similarity threshold $\alpha$ in Eq.~\ref{eq:query_level_aug} is empirically set to 0.6. 
The system-level positive sample expansion module follows the same design. 
For world knowledge enrichment, we utilize the Qwen2.5-14B~\cite{qwen2025qwen25technicalreport} as a pseudo-retriever to synthesize external positive examples. We generate 35 million synthetic positive samples.
During Training, all retrieval models, including our proposed method and the baselines, adopt a dual-encoder architecture. We initialize both the query and document encoders using the Qwen2.5-0.5B pre-trained language model. The query encoder takes the raw query text as input, while the document encoder is fed with the textual content of the photo. The maximum input sequence length is limited to 128 tokens. We set the temperature factor $\tau$ in Eq.~\ref{eq:eq:h-infonce} as a learnable parameter with a initial value of 0.05. 
We train all models using the Adam optimizer~\cite{kingma2015adam}, with a learning rate of 2e-5 and a weight decay of 1e-4.

\subsection{Main Results}
To comprehensively evaluate the effectiveness of CroPS, we conduct experiments on the CPSQA dataset and report three key metrics: NDCG@4, and Recall@100 on CT and QR.

\begin{table}[t]
\centering
\small
\renewcommand{\arraystretch}{1.1}
\begin{tabular}{l*{3}{c}}
\toprule 
\multirow{2}{*}[-0.75ex]{\textbf{Method}} & \multicolumn{2}{c}{\textbf{Recall@100} $\uparrow$ } & \multirow{2}{*}[-0.75ex]{\textbf{\makecell{NDCG@4 $\uparrow$ \\ (\%) }}} \\
\cmidrule{2-3}
 & \textbf{CT(\%)} & \textbf{QR(\%)} & \\
\midrule 
BM25            & 42.9 & 22.5 & 64.8 \\
NCE           & 53.6 & 27.5 & 65.4 \\
DPR  & 56.0 & 30.7 & 66.5 \\
ANCE                & 56.9 & 31.3 & 67.1 \\
ADORE + STAR           & 59.4 & 31.9 & \textbf{67.4} \\
TriSampler  & 59.8 & 32.2 & 66.9 \\
FS-LR   & 59.6 & 33.0 & 66.0 \\
\midrule 
\textbf{CroPS (Ours)}       & \textbf{69.1} & \textbf{40.1} & {67.0} \\
\bottomrule 
\end{tabular}
\caption{Performance comparison of different methods on search ranking tasks.}
\label{tab:results}
\end{table}

The comparison methods fall into three major categories. 
Classical methods: BM25~\cite{robertson2009probabilistic} as a probabilistic ranking baseline and NCE~\cite{gutmann2010noise} as a traditional contrastive learning approach.
Neural network methods: DPR~\cite{karpukhin2020dense} with dual-encoders, ANCE~\cite{xiong2021approximate} using dynamic hard negative sampling, and ADORE+STAR~\cite{zhan2021optimizing} for stable optimization.
Negative sampling strategies: TriSampler~\cite{yang2024trisampler} with principled sampling and FS-LR~\cite{zheng2024full}, which introduces multi-level negative labels.

As shown in Table~\ref{tab:results}, CroPS demonstrates significant advantages in retrieval performance that directly impact user experience.
On the CT dataset, CroPS achieves a Recall@100 of 69.1\%, surpassing the best baseline (TriSampler) by 9.3\%, demonstrating advanced capability in obtaining content that corresponds with user search intent. More importantly, CroPS obtains 40.1\% Recall@100 on QR, which significantly outperforms existing models. It means that the users can find relevant content in their first search attempt, reducing query refinement needs and improving search satisfaction. This demonstrates CroPS's ability to break through information silos and understand user search intent even with imprecise initial queries. 
While NDCG@4 measures ranking quality based on human-annotated relevance judgments, CroPS achieves competitive performance at 67.0\%, remaining on par with top-performing baselines. The key advantage lies in combining extraordinary recall capabilities with maintained ranking precision. This means that with comparable ranking quality, users can immediately find exactly what they're looking for without multiple search iterations. The results establish CroPS as setting new standards for industrial search by delivering both comprehensive content coverage and superior user experience through its multi-source positive sample strategy.

\subsection{Ablation Study}

We conduct ablation studies focusing on core components of CroPS: Data Engine, HLA, and H-InfoNCE Optimization. 

\noindent\textbf{Effectiveness of the CroPS Data Engine.}
We progressively incorporate three augmentation strategies—query-level augmentation, system-level augmentation, and world knowledge enrichment—on top of the baseline that only uses online search data. 
As shown in Table~\ref{tab:data_ablation}, each augmentation consistently improves performance. 
Adding query-level augmentation improves Recall@100 by +3.7\% on CT and notably +3.1\% on QR, showing the value of capturing user intent continuity through query reformulation behavior, with the QR improvement being the largest single-strategy gain across all individual augmentations. 
In addition, including system-level enhancement brings a greater gain of +2.4\% in CT and +1.3\% in QR, confirming the benefit of leveraging recommendation signals to enrich exposure diversity.
Incorporating world knowledge augmentation using LLMs results in the further gain, achieving a final Recall@100 of 69.1\% (CT) and 40.1\% (QR), as well as the best NDCG@4 score of 67.0\%. This shows that LLM-generated examples complement the in-platform data by injecting external semantics and broader world associations. 
These results validate that each augmentation source addresses a unique dimension of the filter bubble, and their combination yields additive gains in both recall and ranking performance.

\begin{table}[t]
    \centering
    \small
    \setlength{\tabcolsep}{1.2mm}
    \renewcommand{\arraystretch}{1.1}
    \begin{tabular}{lccc}
        \toprule
        \multirow{2}{*}[-0.75ex]{\textbf{Model}} & \multicolumn{2}{c}{\textbf{Recall@100} $\uparrow$ } & \multirow{2}{*}[-0.75ex]{\textbf{\makecell{NDCG@4 $\uparrow$\\(\%)} }} \\
        \cmidrule{2-3}
        & \textbf{CT(\%)} & \textbf{QR(\%)} & \\
        \midrule
        \multicolumn{4}{l}{(1) \textit{CroPS Data Engine Ablation}} \\
        \midrule
        Baseline    & 59.6    & 33.0 & 66.0 \\
        ~ + Query-level Aug.    & 63.3 & 36.1    & 66.5    \\
        ~ + System-level Aug. & 65.7    & 37.4  & 66.7    \\
        ~ + World Knowledge Aug. & \textbf{69.1}    & \textbf{40.1}    & \textbf{67.0}   \\
        \midrule
        \multicolumn{4}{l}{(2) \textit{Hierarchical Label Assignment Ablation}} \\
        \midrule
        CroPS$^{\dag}$ {\scriptsize (binary label)}  & 59.9    & 32.1    & 66.7    \\
        CroPS$^{\ddag}$ {\scriptsize (query-level aug. label = 4)} & 67.1    & 38.4    & 66.8     \\
        \bottomrule
    \end{tabular}
    \caption{Data augmentation and HLA ablations.}
    \label{tab:data_ablation}
\end{table}

\noindent \textbf{Effectiveness of Hierarchical Label Assignment (HLA).}
To evaluate the impact of our hierarchical labeling strategy, we conduct two ablation studies as shown in the second part of Table~\ref{tab:data_ablation}. First, we compare the CroPS model with a variant CroPS$^\dag$, where hierarchical labels are simplified into binary form: samples with label $\geq$ 4 are treated as positives, and others as negatives. This variant uses a standard InfoNCE loss for optimization. The results show a significant performance drop, with Recall@100 decreasing by 9.2\% on CT and 8.0\% on QR compared to the CroPS. This highlights the importance of preserving label granularity, as HLA provides more informative supervision that helps the model distinguish varying degrees of relevance. 
We further investigate the effect of assigning the highest label (\textit{i.e.}, 5) to the query-level augmented positives. As discussed in HLA, query-level augmented positives receive the highest label because they represent the most authentic manifestation of user preferences, as these reformulation behaviors inherently encode what users truly seek after experiencing initial search outcomes. To validate this design choice, we modify this setting in CroPS$^\ddag$ by assigning a lower label of 4 to such positives. This change leads to noticeable degradation on Recall@100 by 2.0\% on CT and 1.7\% on QR, confirming that query-level reformulation signals are indeed particularly strong indicators of retrieval relevance and should be weighted accordingly in training. Together, these results demonstrate that HLA offers finer-grained supervision, which significantly contributes to CroPS's retrieval effectiveness.

\begin{table}[t]
    \centering
    \small
    \renewcommand{\arraystretch}{1.1}
    \begin{tabular}{lcccc}
        \toprule
        \multirow{2}{*}[-0.75ex]{\textbf{Loss}} & \multicolumn{2}{c}{\textbf{Recall@100} $\uparrow$ } & \multirow{2}{*}[-0.75ex]{\textbf{\makecell{NDCG@4 $\uparrow$ \\ (\%)}}} & \multirow{2}{*}[-0.75ex]{\textbf{Speed} $\downarrow$}\\
        \cmidrule{2-3}
        & \textbf{CT(\%)} & \textbf{QR(\%)} & \\
        \midrule
        InfoNCE & 67.8    & 38.9    & 66.9  & 178h \\
        Softmax-CE  & 65.3    & 37.8    & 66.8 &  89h \\
        \textbf{H-InfoNCE}  & {\bf 69.1}    & {\bf 40.1}    & {\bf 67.0}  & {\bf 88h}   \\
        \bottomrule
    \end{tabular}
    \caption{Contrastive optimization comparison.}
    \label{tab:loss_ablation}
\end{table}

\noindent \textbf{Efficiency of H-InfoNCE.}
We compare H-InfoNCE loss against standard InfoNCE and Softmax cross-entropy~\cite{magnani2022semantic} losses. The results are shown in Table~\ref{tab:loss_ablation}. 
InfoNCE compares each positive sample with its corresponding negative instances within a single training step. It leads to the slowest training efficiency.  
Although Softmax-CE trains faster, it yields the worst results. The high-level samples are mistakenly introduced when computing the low-level contrastive loss, even with score regularization to address this issue in Softmax-CE. 
In contrast, H-InfoNCE not only achieves the best performance across all metrics, but also maintains high training efficiency. By computing multi-level contrastive loss in a single forward pass, it significantly improves both accuracy and efficiency. Such efficiency is particularly beneficial in industrial scenarios, where models must be trained frequently on large-scale data with limited computational resources.

\subsection{Online Testing}
We validate the proposed CroPS method with online A/B testing on Kuaishou Search. 

\noindent {\bf Scalability.} Our CroPS is agnostic to the model architecture. We evaluate our method on both dense and sparse model. The dense model is the previously mentioned Qwen2.5-0.5B, which predominantly processes text input. The sparse model employs the MLP structure and utilizes ID features as input. Consistent improvements are observed in Table \ref{tabel:recall_ratio} and Table \ref{tabel:ab_res}, when applying CroPS to these two distinct model types, demonstrating the scalability of our approach. 

\noindent {\bf Recall Capability.} We investigate how our recall results are exposed in the ranking stage. It includes two metrics: Ratio$_\text{rank}$ and Ratio$_\text{show}$. Ratio$_\text{rank}$ represents the ratio of our recall set within the ranking set, while Ratio$_\text{show}$ indicates the ratio of our recall set within the show set (showing to users). As shown in Table \ref{tabel:recall_ratio}, the CroPS on dense model contributes to a 11.1\% increase in Ratio$_\text{rank}$, a 11.8\% increase in Ratio$_\text{show}$. The higher recall ratio reflects better retrieval of relevant content.

\begin{table}[t] 
\centering
\small
\renewcommand{\arraystretch}{1.1}
\begin{tabular}{lccc} %
\toprule 
\textbf{Model Type} & \textbf{Model} & \textbf{Ratio}$_\text{rank}$ $\uparrow$  & \textbf{Ratio}$_\text{show}$ $\uparrow$  \\
\midrule  
\multirow{2}{*}{Dense Model} & Baseline  & 29.8\textsl{\%} & 32.5\textsl{\%} \\
 & CroPS & \textbf{40.9\textsl{\%}} &  \textbf{44.3\textsl{\%}}  \\
\midrule  
\multirow{2}{*}{Sparse Model} & Baseline  & 20.8\textsl{\%} & 29.1\textsl{\%} \\
 & CroPS & \textbf{33.8\textsl{\%}} &  \textbf{44.7\textsl{\%}}  \\
\bottomrule  
\end{tabular}  
\caption{The improvements of CroPS method in Ratio$_\text{rank}$ and Ratio$_\text{show}$ compared to the production baseline.}
\label{tabel:recall_ratio} 
\end{table}

\begin{table}[t] 
\centering
\small
\renewcommand{\arraystretch}{1.1}
\begin{tabular}{cccc} %
\toprule 
\textbf{Model Type} & \textbf{CTR} $\uparrow$ & \textbf{LTR} $\uparrow$ & \textbf{RQR} $\downarrow$ \\
\midrule  
Dense Model & +0.869\textsl{\%} & +0.483\textsl{\%}  &  -0.646\textsl{\%}  \\
Sparse Model & +0.783\textsl{\%} & +0.423\textsl{\%}  &  -0.614\textsl{\%}  \\
\bottomrule  
\end{tabular}  
\caption{The improvements of CroPS in online A/B test.}
\label{tabel:ab_res} 
\end{table} 

\noindent {\bf User Satisfaction.} To further evaluate the user's interaction with the search results page after searching, we introduce the metrics of CTR, LPR (long-play rate), and RQR (reformulated-query rate). CTR assesses whether the search results meet user needs and drive click behavior, LPR measures the proportion of users who watch a video for an extended duration, and RQR is the proportion of users who change their query and initiate a search again. The results of CroPS are shown in Table \ref{tabel:ab_res}, our comprehensive CroPS method on dense model contributes to a 0.869\textsl{\%} increase in CTR, a 0.483\textsl{\%} increase in LPR and a 0.646\textsl{\%} decrease in the RQR. The observed high click-through and long watch times, coupled with a low rate of query reformulation, indicate that users successfully located their intended content. This demonstrates that our model effectively captures and fulfills user search intent.

\begin{figure}[t]
    \centering 
    \includegraphics[width=0.45\textwidth]{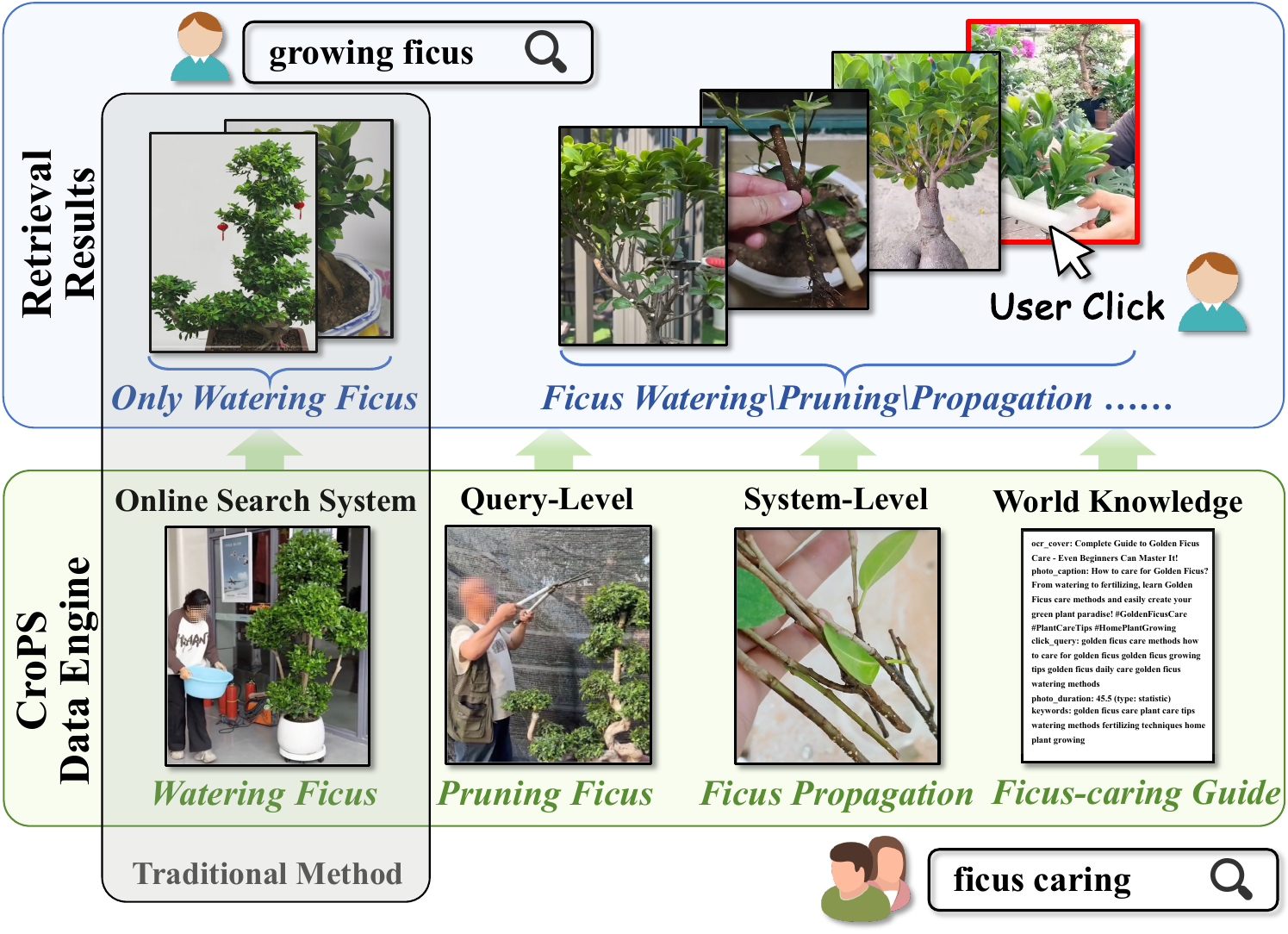}
    \caption{Case Study. Blue block indicates test results, while green block shows evidence traces from our method.}
    \label{fig:example} 
\end{figure}

\subsection{Case Study}
Considering query ``growing ficus'', as shown in the blue block of Figure \ref{fig:example}, traditional methods retrieve only watering or fertilizing videos, while CroPS retrieves diverse content including propagation, pruning, and disease prevention videos. The user ultimately clicks on the propagation video from our results. As shown in the green block of Figure \ref{fig:example}, we further explore the supporting evidence of the CroPS data engine by retrieving similar queries from the training data using the original query as an anchor. We observe that traditional click-based training data primarily consists of basic care videos.  In contrast, CroPS's data engine enriches training samples with diverse content like propagation techniques, pruning guides, and pest control from multiple channels, capturing various user intents beyond clicked results. This design breaks information bubbles by expanding training diversity, better satisfying varied user needs.

\section{Conclusion}

To address information cocoons in industrial dense retrieval systems, we propose CroPS, which enriches the positive training signal with multiple perspectives, offering relevant yet novel supervision. Hierarchical Label Assignment and H-InfoNCE enhance fine-grained retrieval semantics. Future work will integrate CroPS with generative retrieval methods.

\bibliography{aaai2026}

\newpage
\appendix

\lstset{
  language=Python,        
  basicstyle=\ttfamily\small,  
  numbers=left,           
  numberstyle=\tiny\color{gray}, 
  keywordstyle=\color{blue},    
  commentstyle=\color{gray},   
  stringstyle=\color{black},     
  breaklines=true,        
  frame=leftline,           
  captionpos=b            
}

\section{Appendix}
\subsection{H-InfoNCE Implementation}

\begin{figure*}[t]
\centering
\includegraphics[width=1.0\textwidth]{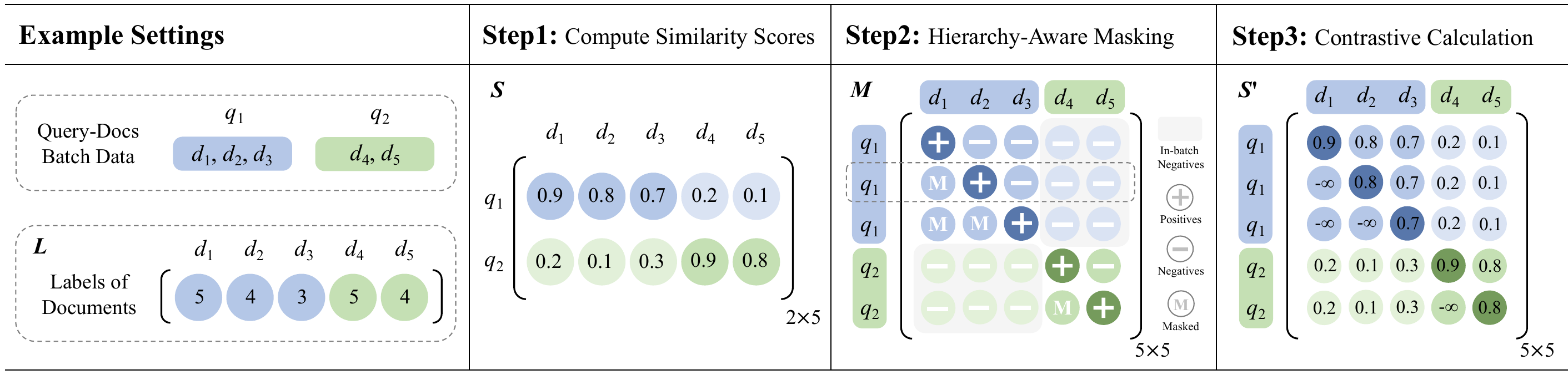}
\caption{The algorithmic process of H-InfoNCE. As an example, we use training data consisting of 2 queries ($q_1$ and $q_2$ with their corresponding documents: $d_1,d_2,d_3$ for $q_1$, and $d_4,d_5$ for $q_2$, respectively) for a single forward pass. It can primarily be divided into three steps: Compute Similarity Scores, Hierarchy-Aware Masking and Contrastive Calculation.}
\label{fig:hinfonce}
\end{figure*}

We develop an efficient algorithm for computing the H-InfoNCE loss by leveraging a carefully designed data organization strategy. Specifically, video samples across different queries are processed in a unified format. 
To elucidate our algorithm, we illustrate the detailed process of H-InfoNCE with a concrete example in Figure~\ref{fig:hinfonce}. The example utilizes training data consisting of 2 queries ($q_1$ and $q_2$ with their corresponding documents: $d_1$, $d_2$, $d_3$ for $q_1$, and $d_4$, $d_5$ for $q_2$, respectively) for a single forward pass, along with their hierarchical label assignments \textbf{\(L\)}. The procedure is composed of three main steps. 

\noindent\textbf{Step 1: Compute Similarity Scores.} The process begins by computing an initial similarity score matrix \(S\), between all query embeddings (\(q_1, q_2\)) and all document embeddings (\(d_1, \dots, d_5\)) within the batch.

\noindent\textbf{Step 2: Hierarchy-Aware Masking.} The core of our method lies in the construction of a hierarchy-aware mask $M$. This mask clearly establishes the contrastive relationship among documents for a given query. For each query, the labeling process for the document is as follows:
\begin{itemize}
    \item \textbf{Positive (+):} Within the query, each document is treated as an individual positive sample. For instance, $d_2$ serves as the positive sample for $q_1$.
    \item \textbf{Negative (-):} Negatives include two types: (1) all documents associated with other different queries (\textit{i.e.}, standard in-batch negatives) and (2) documents belonging to the same query but possessing a strictly lower label in $L$.
    \item \textbf{Masked (M):} Crucially, any document from the same query that has a higher or equal label is masked out. For example, when \(d_2\) (label 4) is the positive sample, \(d_1\) (label 5) is masked rather than treated as negative sample.
\end{itemize}

\noindent\textbf{Step 3: Contrastive Calculation.} Finally, we compute the logits for the loss function. The initial similarity matrix \(S\) is first expanded into a \(N \times N\) ($N$ is the number of all documents, $N=5$ in Figure \ref{fig:hinfonce}) matrix, where each row contains the similarity scores from its corresponding query. The mask \(M\) is then applied to this matrix, setting the scores of all masked pairs to \(-\infty\). This results in the final logit matrix \(S'\), which is subsequently used to calculate the H-InfoNCE loss according to Eq.~\ref{eq:eq:h-infonce}.

The efficient implementation of H-InfoNCE ensures high scalability. As evidenced by our experiments, it achieves comparable computational speed to standard InfoNCE while leveraging multi-level supervision. This property is crucial for industrial applications, where models are frequently trained on massive datasets under strict time and resource constraints.

\subsection{Dataset, Metrics and Implementation Details}

\noindent {\bf Dataset.} 
Considering data security and user privacy concerns, we are unable to make our entire dataset publicly available. The dataset statistics are presented in Table~\ref{tabel:appdenix_statistics}. To further enhance reproducibility, we provide the example of the CPSQA training set, as shown in Listing~\ref{lst:data}.

\null
\begin{lstlisting}[label={lst:data}, language=Python, caption=The data example from CPSQA.]
photo_item = {
    "ocr_cover": "...",
    "photo_caption": "...",
    "keyword": "...",
    "click_query": "...",
    "play_cnt": 2393,
    "like_cnt": 798,
    "long_view_cnt": 125,
    "download_cnt": 87
}

data_item = {
    "user_id": 34758937,
    "query": "transformer",
    "photo_list": [
        {
            "photo_id": 1938937936,
            "photo_content": photo_item,
            "hierarchical_label": 4
        },
        ...,
        {
            "photo_id": 1037689405,
            "photo_content": photo_item,
            "hierarchical_label": 2
        }
    ]
}

training_data = [
    data_item_1, 
    data_item_2, 
    ..., 
    data_item_N
]
\end{lstlisting}

\noindent {\bf Metrics.} 
We assess our retrieval system employing two standard metrics: Recall and Normalized Discounted Cumulative Gain (NDCG).
Recall@K measures the proportion of relevant documents retrieved from the ground truth within the top-K results. For a query $q$ with a ground truth set $R_q$ and a retrieved set $S_q^{(K)}$ (top-K results), Recall@K is computed as:
\begin{equation}
\text{Recall@K} = \frac{1}{|Q|}\sum_{q\in Q} \frac{|R_q \cap S_q^{(K)}|}{|R_q|}
\label{eq:eq:recall_k}
\end{equation}
where $|Q|$ is the total number of queries. In our experiments, we report Recall@100. 
NDCG@K measures the model’s ability to rank the top K results.  
The NDCG is start with Discounted Cumulative Gain (DCG) at K:
\begin{equation}
\text{DCG@K} = \sum_{i=1}^{K} \frac{rel_i}{\log_2(i+1)}
\label{eq:eq:dcg}
\end{equation}
where $rel_i = 2^{\text{label}} - 1$ as the graded relevance score at position $i$. Then, Ideal DCG (IDCG@K) is the maximum possible DCG@K for a perfectly ranked list. Finally, Normalized DCG (NDCG) is calculated as:
\begin{equation}
\text{NDCG@K} = \frac{\text{DCG@K}}{\text{IDCG@K}}
\label{eq:eq:Ndcg}
\end{equation}
In our study, we calculate NDCG@4 as a reference monitoring metric.

\begin{table}[t] 
\centering
\small
\renewcommand{\arraystretch}{1.1}
\resizebox{1\columnwidth}{!}{
\begin{tabular}{c|cccc} %
\toprule 
\textbf{Split Type} & \textbf{Training} & \textbf{CT} & \textbf{QR} &  \textbf{Annotation} \\
\midrule  
\textbf{Number of Samples} & 500M & 10K  &  10K & 0.5K  \\
\bottomrule  
\end{tabular}  
}
\caption{The statistics of CPSQA dataset, where ``K'' and ``M'' indicate thousand and million respectively. Both CT and QR are constructed to reflect Recall@100, while the Annotation set is manually labeled for evaluating NDCG@4.}
\label{tabel:appdenix_statistics} 
\end{table} 

\begin{figure}[t]
    \centering 
    \includegraphics[width=0.45\textwidth]{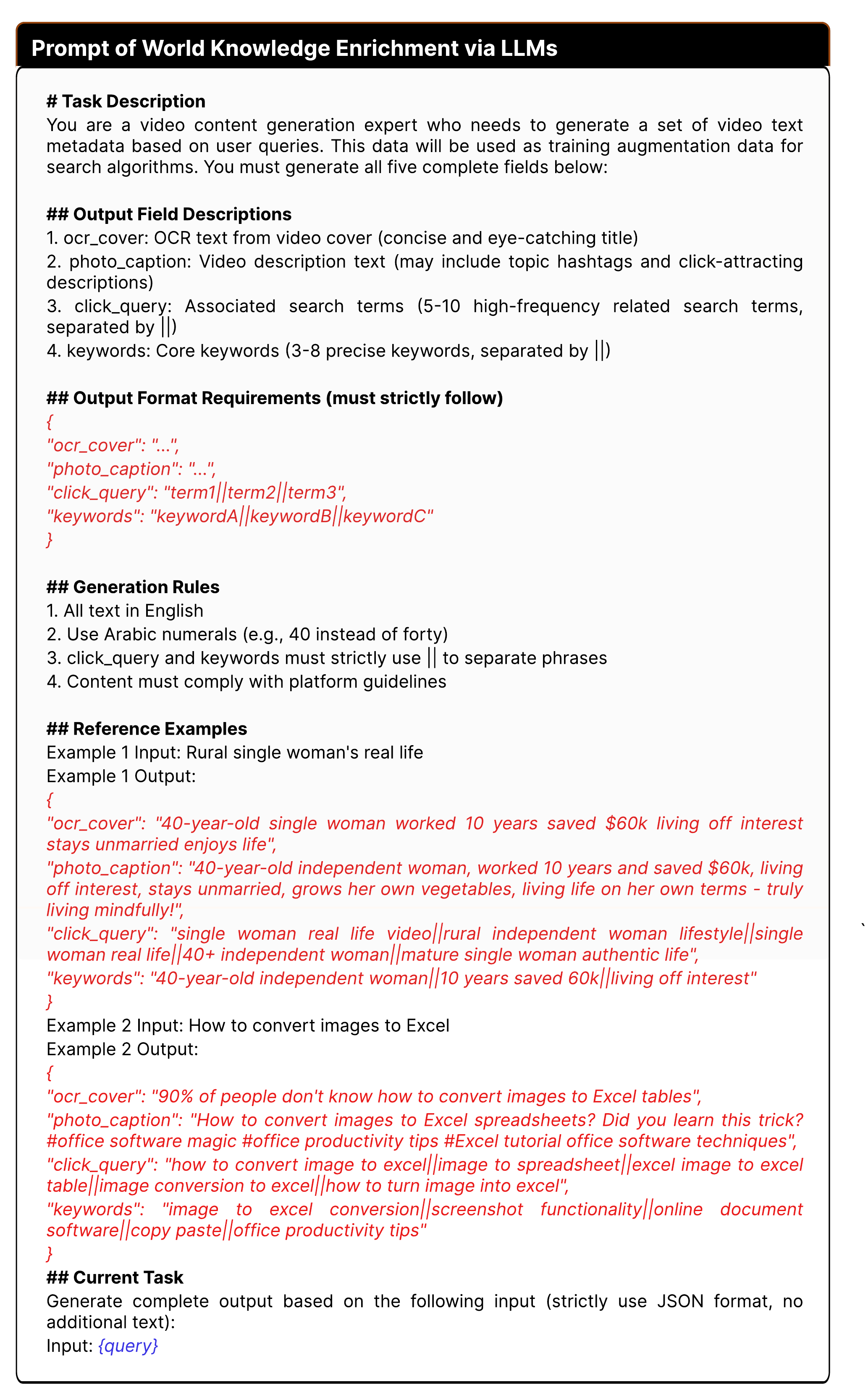}
    \caption{LLM Prompt in World Knowledge Enrichment.} 
    \label{fig:Prompt} 
\end{figure}

\noindent {\bf Implementation Details.}  
As illustrated in Figure~\ref{fig:Prompt}, we employ a one-shot prompting strategy to generate synthetic training data. This process involves providing a Large Language Model (LLM) with a detailed prompt that includes a user's query and a single, high-quality exemplar. This exemplar serves as the ``one-shot'' example, demonstrating the desired output format and style. The LLM is explicitly instructed to generate a structured set of video metadata, comprising four key fields: ocr cover (the video's title), photo caption (the descriptive text), click query (a list of potential user search terms), and keywords (core thematic tags). The prompt enforces strict formatting requirements, such as mandating a JSON output and the use of specific delimiters ($\Vert$) for search terms and keywords, ensuring the generated data is machine-readable and ready for downstream processing.

\subsection{Extended Online Testing}

\begin{table}[t] 
\centering
\caption{The improvements of CroPS method in Ratio$_\text{rank}$ and Ratio$_\text{show}$ compared to the production baseline across both sparse and dense models.}
\label{tabel:Appendix_recall_ratio} 
\renewcommand{\arraystretch}{1.3}
\resizebox{1.0\columnwidth}{!}{
\begin{tabular}{llcc}
  
\toprule
\textbf{Model Type} & \textbf{Model} & \textbf{Ratio}$_\text{rank}$ $\uparrow$  & \textbf{Ratio}$_\text{show}$ $\uparrow$  \\
\midrule 
\multirow{4}{*}{Dense Model} & Baseline  & 29.8\textsl{\%} & 32.5\textsl{\%} \\
 &  + Query-level Aug.  & 34.8\textsl{\%} & 35.1\textsl{\%} \\
 & + System-level Aug. & 38.7\textsl{\%} & 41.8\textsl{\%}   \\
 & + World Knowledge Aug. & \textbf{40.9\textsl{\%}} &  \textbf{44.3\textsl{\%}}  \\
\midrule
\multirow{4}{*}{Sparse Model} & Baseline  & 20.8\textsl{\%} & 29.1\textsl{\%} \\
 & + Query-level Aug  & 27.9\textsl{\%} & 38.7\textsl{\%} \\
 & + System-level Aug. & 28.6\textsl{\%} & 40.6\textsl{\%}   \\
 & + World Knowledge Aug. & \textbf{33.8\textsl{\%}} &  \textbf{44.7\textsl{\%}}  \\
\bottomrule
\end{tabular}  
}
\end{table}

\begin{table}[t] 
\centering
\caption{The improvements of CroPS method in online A/B test compared to the production baseline across both sparse and dense models.}
\label{tabel:Appendix_ab_res} 
\renewcommand{\arraystretch}{1.3}
\resizebox{1.0\columnwidth}{!}{
\begin{tabular}{llccc} 
  
\toprule 
\textbf{Model Type} & \textbf{Model} & \textbf{CTR} $\uparrow$ & \textbf{LTR} $\uparrow$ & \textbf{RQR} $\downarrow$ \\
\midrule 
\multirow{3}{*}{Dense Model} & + Query-level Aug.  & +0.448\textsl{\%} & +0.222\textsl{\%} & -0.334\textsl{\%} \\
 & + System-level Aug. & +0.756\textsl{\%} & +0.361\textsl{\%} & -0.487\textsl{\%}   \\
 & + World Knowledge Aug. & \textbf{+0.869\textsl{\%}} & \textbf{+0.483\textsl{\%}}  &  \textbf{-0.646\textsl{\%}}  \\
\midrule   
\multirow{3}{*}{Sparse Model} & + Query-level Aug.  & +0.328\textsl{\%} & +0.187\textsl{\%} & -0.297\textsl{\%} \\
& + System-level Aug. & +0.573\textsl{\%} & +0.310\textsl{\%} & -0.458\textsl{\%}   \\
& + World Knowledge Aug. & \textbf{+0.783\textsl{\%}} & \textbf{+0.423\textsl{\%}}  &  \textbf{-0.614\textsl{\%}}  \\
\bottomrule   
\end{tabular}  
}
\end{table} 

To verify the impact of different positive samples in CroPS, we conduct online A/B testing with various positive samples in Kuaishou Search. Both the experimental group and control group are randomly assigned 10\textsl{\%} of online users for a 7-day A/B test. As shown in Table \ref{tabel:Appendix_recall_ratio} and Table \ref{tabel:Appendix_ab_res}, model performance is improved by the inclusion of diverse positive samples.

\noindent {\bf Scalability.} As shown in Table \ref{tabel:Appendix_recall_ratio} and Table \ref{tabel:Appendix_ab_res}, each augmentation consistently improves performance in both dense model and sparse model. This result highlights the scalability of our approach. Since the sparse and dense models yield similar conclusions, we mainly report results for the dense model below.

\noindent {\bf Recall Capability.} 
As shown in Table \ref{tabel:Appendix_recall_ratio}, adding query-level augmentation to the dense model improves Ratio$_\text{rank}$ by +5\%  and Ratio$_\text{show}$ by +2.6\%, showing the value of capturing user intent continuity through query reformulation behavior. In addition, including system-level enhancement in the dense model brings a gain of +3.9\% in Ratio$_\text{rank}$ and +6.7\% in Ratio$_\text{show}$, confirming the benefit of leveraging recommendation signals to enrich exposure diversity. In the context of the dense model, incorporating world knowledge augmentation using LLMs results in the further gain, achieving a final Ratio$_\text{show}$ of 40.9\% and Ratio$_\text{show}$ of 44.3\%. This demonstrates that each positive example can enhance the model's ability to recall relevant videos.

\noindent {\bf User Satisfaction.}
As shown in Table \ref{tabel:Appendix_ab_res}, with the addition of query-level positives to the baseline, the dense model contributes to a 0.448\textsl{\%} increase in CTR, a 0.222\textsl{\%} increase in LPR and a 0.334\textsl{\%} decrease in the RQR, showing the value of capturing user intent continuity through query reformulation behavior. Notably, query-level positives are the most effective in reducing the RQR for both dense and sparse models. In addition, including system-level enhancement in the dense model brings a 0.308\textsl{\%} increase in CTR, a 0.139\textsl{\%} increase in LPR and a 0.153\textsl{\%} decrease in the RQR, confirming the benefit of leveraging recommendation signals to enrich exposure diversity. Incorporating world knowledge augmentation using LLMs results in the further gain, the dense model contributes to a 0.113\textsl{\%} increase in CTR, a 0.122\textsl{\%} increase in LPR and a 0.159\textsl{\%} decrease in the RQR. This indicates that each positive example is useful for capturing the user search intent.

Our method shows promising results, demonstrating the effectiveness of positive samples from multiple perspectives. This approach is particularly beneficial in industry, where positive samples from multiple perspectives may be rich, enabling faster improvements in retrieval performance.

\subsection{Further Analysis}

\noindent We analyze the positive examples generated by the LLM as well as the learnable temperature coefficient factor $\tau$.

\begin{figure}[t]
    \centering 
    \includegraphics[width=0.4\textwidth]{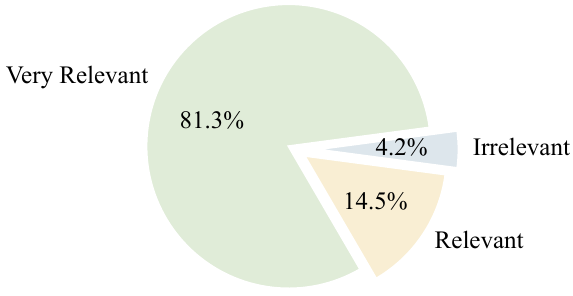}
    \caption{Manual evaluation of LLM-generated positives.} 
    \label{fig:pie} 
\end{figure}

\noindent \textbf{LLM-Generated Positives.} 
As detailed in our Methodology, we adopt a standard industry practice of using textual information as the primary video representation. This strategic choice enables the direct use of LLMs to generate synthetic video descriptions for data augmentation.
To evaluate the relevance of the LLM-generated samples, we conduct a rigorous manual evaluation. A random sample of 1,000 generations is assessed on a three-point relevance scale (Very Relevant, Relevant, Irrelevant). To ensure consistency in our manual evaluation, the three-point scale was defined with specific criteria. ``Very Relevant'' generations are those that accurately and comprehensively address the search terms. ``Relevant'' samples somewhat address the search terms but may not cover all aspects comprehensively. Finally, ``Irrelevant'' samples are those that do not adequately address the search terms or are off-topic. The results shown in Figure~\ref{fig:pie} confirm the high quality of our approach: 81.3\% are rated ``Very Relevant'', 14.5\% ``Relevant'', and only 4.2\% ``Irrelevant''. These results indicate a strong performance of the LLM in generating content that is closely aligned with user queries, with 95.8\% of the outputs being either ``Very Relevant'' or ``Relevant''. This high level of relevance reinforces the effectiveness of using LLMs as a pseudo-retriever for enriching the training dataset. The primary goal is to generate new video descriptions that are conceptually aligned with the input query. These synthetically generated descriptions act as positive samples, significantly enriching our training dataset. This method allows us to augment our data with a wider range of diverse linguistic expressions, creative conceptual variations, and external world knowledge that may not be present in the platform’s existing content or historical user search logs.

\begin{figure}[t]
    \centering 
    \includegraphics[width=0.4\textwidth]{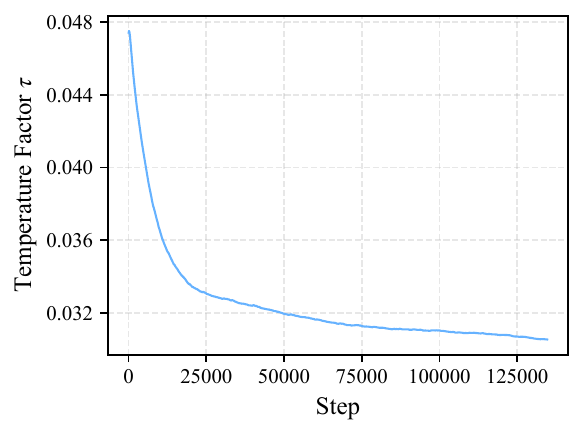}
    \caption{Dynamic adjustment of the temperature factor $\tau$.} 
    \label{fig:temperature} 
\end{figure}

\noindent \textbf{Temperature Factor.} In this part, we analyze how the learnable $\tau$ in Eq.~\ref{eq:eq:h-infonce} changes as the training steps advance. As shown in Figure \ref{fig:temperature}, $\tau$ gradually decreases during training and eventually converges around 0.03. This dynamic adjustment reveals an intrinsic mechanism of the model's learning process. A lower temperature sharpens the softmax distribution, increasing the penalty for misdistinguishing the positive sample from hard negatives. Initially, a higher $\tau$ allows the model to explore on a smoother loss landscape, preventing instability from an overly difficult task. As the model's representation ability improves, it automatically lowers $\tau$, effectively increasing the contrastive learning difficulty to compel itself to learn more discriminative, fine-grained features. Its convergence to a stable value indicates that the model has found an optimal balance between effective sample discrimination and training stability. This self-adaptive parameter tuning not only obviates manual hyperparameter tuning but, more importantly, reflects the model's learning progression from coarse-grained perception to fine-grained discrimination.

\subsection{Additional Case Study}

\begin{figure*}[t]
    \centering 
    \includegraphics[width=1\textwidth]{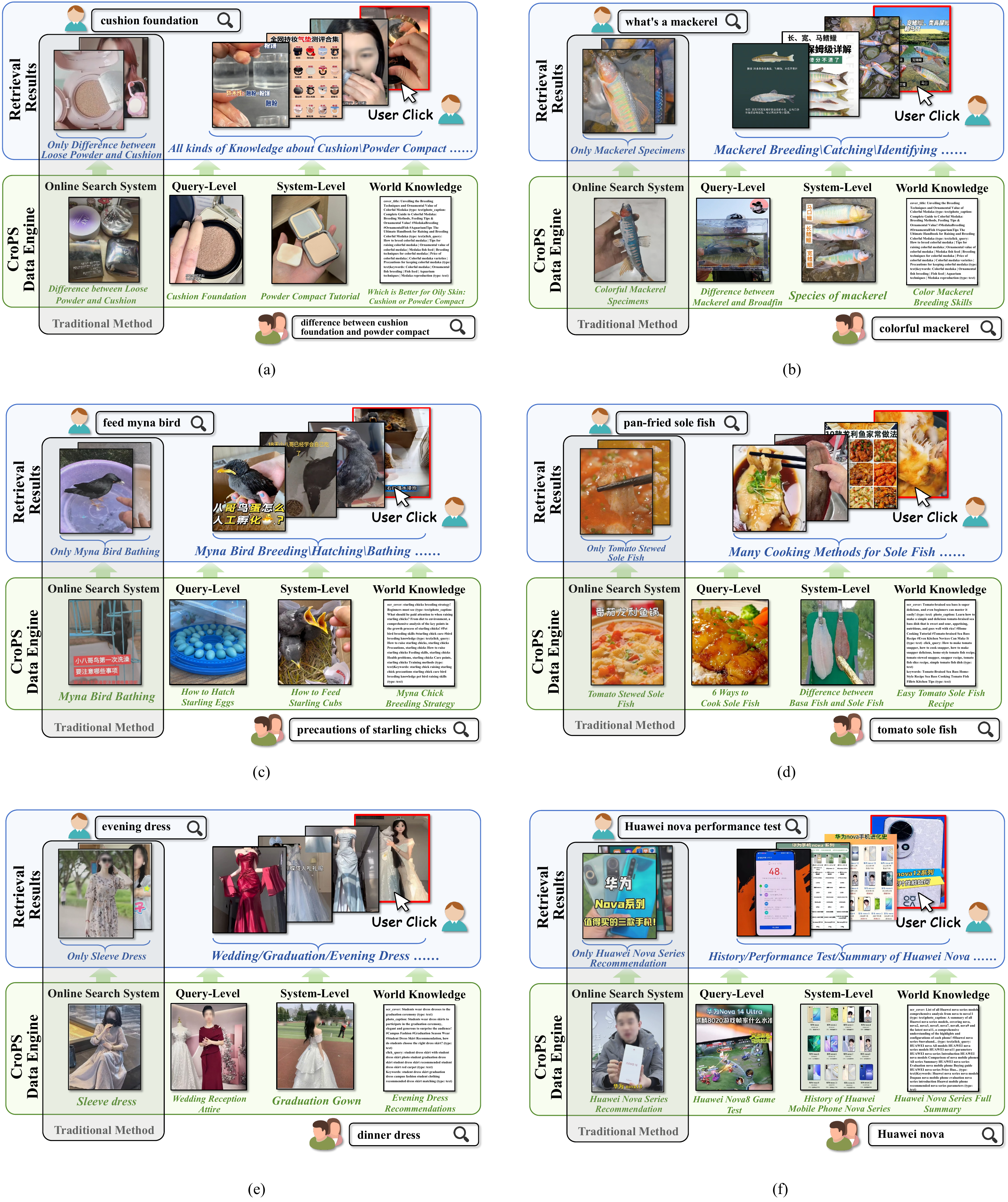}
    \caption{Case Study. Blue block indicates test results, while green block represents the evidence traces found through our method.}
    \label{fig:example2}
\end{figure*}

As illustrated in Figure~\ref{fig:example2}, six additional cases are provided. We take Figure~\ref{fig:example2}.a as an example for analysis. Considering query ``What's the difference between cushion foundation and powder compact'', traditional retrieval using only clicked videos from online search systems yields limited results focusing on ``difference between loose powder and cushion'', missing the direct comparison users seek. In contrast, CroPS retrieves comprehensive content including practical application tutorials, skin type recommendations, and direct product comparisons. The user ultimately clicks on the comparison video addressing specific skin concerns from our results. We further explore the supporting evidence of the CroPS data engine by retrieving similar queries from the training data using the original query as an anchor. We observe that traditional click-based training data primarily consists of generic makeup tutorials. CroPS's data engine enriches training samples with diverse content like usage tutorials, skin type comparisons, and product recommendations from multiple channels, capturing various user intents beyond clicked results. This multi-perspective approach breaks the filter bubble by expanding training diversity, ultimately improving both retrieval diversity and relevance. The same phenomenon is consistently observed in the remaining cases presented in Figure~\ref{fig:example2}.

\end{document}